\documentstyle[psfig,epsf,aps,preprint]{revtex}

\def\tr{\mbox{\bf tr }}

\title{\Large Frustrated Blume-Emery-Griffiths model}

\author{ {\large Georg R. Schreiber}\thanks{E-mail: georg@spht.saclay.cea.fr}\\
 {\small Service de Physique Th\'eorique, C.E.A.-Saclay, 
           F - 91191 Gif-sur-Yvette Cedex France} \\
 {\small and Division de Physique Th\'eorique, Institut de Physique 
           Nucl\'eaire} \\
 {\small Universit\'e\ Paris-Sud F--91406 Orsay Cedex France} }

\begin{document}

\maketitle

\vskip 1 cm

\begin{abstract}
  A generalised integer S Ising spin glass model is analysed
  using the replica formalism. The bilinear couplings are
  assumed to have a Gaussian distribution with ferromagnetic
  mean $<J_{ij}> = J_0$. Incorporation of a quadrupolar
  interaction term and a chemical potential leads to a richer
  phase diagram with transitions of first and second order.
  The first order transition may be interpreted as a phase
  separation, and contrary to what has been argued previously,
  it persists in the presence of disorder. Finally, the
  stability of the replica symmetric solution with respect
  to fluctuations in replica space is analysed, and the
  transition lines are obtained both analytically and numerically.
\end{abstract}

\clearpage

\section{Introduction}

  The much studied SK model of spin glasses may be generalised
  in different ways. The standard spin glass model, given by

  \begin{equation}
  H = - \sum_{<ij>} J_{ij} s_i s_j \qquad \mbox{ with } \qquad s_i = \pm 1,
  \end{equation}

  \noindent may be extended for instance by allowing values
  $s = 0, \pm 1, \pm 2, \ldots, \pm S$ for the spin variables.
  It is then possible to consider higher order interactions
  such as $K \sum_{ij} s_i^2 s_j^2$,
  or a chemical potential such as $\Delta \sum_i s_i^2$.
  Such generalisations can be regarded as extensions of the
  Blume-Emery-Griffiths model (BEG) \cite{Blume_Emery_Griffiths_71}.
  Indeed the BEG model allows $s = 0, \pm 1$ and
  takes into account the aforementioned higher
  order interaction and chemical potential. However, in the standard
  formulation, the
  bilinear couplings are neither frustrated nor disordered: they
  are ferromagnetic. The $s = 0$ degree of freedom has been used
  to model a diluted
  lattice gas and leads to a first order transition separating
  magnetic and non-magnetic phases. Thus our generalisation of the
  BEG Hamiltonian is a good way to study the influence of disorder
  and frustration on first order transitions.

  The BEG model has been studied in various contexts and a mean field
  approximation has been given by Blume, Emery and Griffiths
  \cite{Blume_Emery_Griffiths_71}.
  They introduced the model in order to study multicritical
  phenomena associated with physical systems such as binary mixtures.
  For an overview, see the review article
  of Lawrie and Sarbach \cite{Lawrie_Sarbach_83}.
  To improve the mean field
  results and to cope with the finite dimensionality of real physical
  systems, several different techniques have been applied to the
  BEG model, e.g., Tucker
  \cite{Tucker_88} applied the exponential operator technique of Honmura
  and Kaneyoshi \cite{Honmura_Kaneyoshi_79} to the isotropic BEG, and
  Fittipaldi {\em et al.} \cite{Fittipaldi_Kaneyoshi_89} applied it to
  the anisotropic BEG model. In addition the model has been treated in
  the cluster approximation by Tucker \cite{Tucker_Saber_Peliti_94}
  and in the local mean field approximation by Maritan {\em et al.}
  \cite{Buzano_Maritan_Pelizzola_94}.

  Several {\it disordered} BEG models have been studied.
  Berker {\em et al.} \cite{Berker_Falicov_95_b} looked
  at a bond disordered BEG model and Ez-Zahraouy \cite{Ez_Zahraouy_95}
  looked at a bond diluted BEG model. To our knowledge no studies of
  the case of quenched disorder with {\it frustration}
  have been carried out for the BEG model, though Arenzon
  {\em et al.} \cite{Arenzon_Nicodemi_Sellitto_96} have considered
  a frustrated lattice gas model similar to the BEG model.
  The closest model to a generalised BEG 
%  In contrast to the latter work we will study the BEG model in an
%  approximation, where the spin degrees of freedom $\tau_i = \pm 1$
%  are not decoupled from the lattice gas site variables $n_i = 0,1$.
%  This amounts in annealing the gas site variable rather than
%  quenching them; we are only concerned with one spin variable
%  $S_i = n_i \tau_i$. Even though the connection between
%  both models is apparent through the latter transformation,
%  the physics of the two models is rather different.
%  The second generalisation of the SK spin glass Hamiltonian to
%  include integer spin values
  studied in the literature was first analysed by Ghatak and
  Sherrington \cite{Ghatak_Sherrington_77}. In their model, 
  they considered $s = 0, \pm 1$, and the 
  influence of a chemical potential, but they had
  no quadrupolar interaction and the ferromagnetic mean
  $J_0$ of their bilinear coupling $J_{ij}$ was assumed to vanish.
  The interesting point we wish to stress here is that
  the first order transition of the BEG model, which can be interpreted
  as a phase separation transition, persists 
  in the GS generalisation. This is 
%  transition \cite{Ghatak_Sherrington_77} in addition to the already
%  known spin glass transition. 
%  This first order transition seems to persist
%  in disordered generalisations, which is in contrast to the findings
  in contrast to the findings
  of Berker {\em et al.} which will be discussed below.

  In the present paper we generalise the standard SK spin glass
  Hamiltonian to incorporate
  the chemical potential, the quadrupolar interaction,
  randomly distributed bilinear couplings with a non-zero 
  ferromagnetic mean, and 
  other integer values for the spin variables. We restrict
  our study to the $s = 0,\pm 1$ case, which should
  cover in a qualitative way the general integer $S$-spin models
  \cite{Lage_de_Almeida_82}. Furthermore we do not study the
  effect of disorder in the quadrupolar couplings; such effects
  have been considered in other models by Snowman {\em et al.}
  \cite{Snowman_McKay_94}.
%  Unfortunately we are not aware of any physical counterpart for
%  this model even if its parents, the BEG model and the SK model,
%  have well defined physical situations: The SK model describes
%  the situation of magnetic impurities in a host crystal, whereas
%  the BEG model describes a binary mixture (e.g. He$^3$ and He$^4$).
  The paper is organised as follows. In section 2 we introduce the
  model. In section 3 the free energy is derived in the replica
  symmetric approximation and in section 4 we give the model's phase
  diagram. In section 5 we discuss our results in the light
  of recent arguments of Berker {\em et al.} concerning the
  influence of quenched disorder on phase transitions of first
  order. In section 6 we analyse the stability of the replica
  symmetric solution and derive the lines of instability.
%  A first step to reestablish stability is undertaken in section 7
%  using replica symmetry breaking.
  Conclusions are drawn in the final section.

\section{The model}

  We consider the model described by the BEG-SK Hamiltonian, where the
  spin variables are allowed to assume the values $s_i = 0, \pm 1$

  \begin{equation}
  H = - \frac{1}{2}\sum_{<ij>} J_{ij} s_i s_j  + \Delta \sum_i s_i^2
  - \frac{1}{2} K \sum_{<ij>} s_i^2 s_j^2 - h \sum_i s_i.
  \end{equation}

  This model will be treated in the mean field approximation, {\it i.e.},
  in the infinite range limit.
  For the sake of simplicity we assume, as in the SK model,
  bilinear couplings with a Gaussian
  distribution about a non-zero mean $<J_{ij}> = J_0 > 0$
  allowing for ferromagnetic ordering.
  To avoid additional complexity of the model we consider only
  the case of positive quadrupolar coupling, $K>0$, and a
  ferromagnetic mean $J_0$ of the bilinear couplings.
  The chemical potential $\Delta$ is allowed
  to assume positive as well as negative values.

  The couplings must be rescaled for the present case of infinite ranged
  couplings in order to give a sensible free energy in the thermodynamic
  limit: $J \to J/\sqrt{N}$, $J_0 \to J_0/N$ and $K \to K/N$, respectively.

  \[p(J_{ij}) = \sqrt{\frac{N}{2\pi J^2}}
  e^{\frac{N}{2 J^2}(J_{ij} - J_0/N)^2} . \]

  We consider now some special cases, which have already been treated in
  the literature. When $J = 0 $ and $J_0 > 0$, we recover the non-frustrated
  and non-disordered standard BEG model, which describes a binary
  mixture (e.g., He$^3$ and He$^4$). When $ J_0 = 0$ and $K = 0$,
  we recover the Gathak and Sherrington model.
  In order to cope with the problem of averaging over quenched disorder,
  it is necessary to average the free energy over the bond distribution
  $J_{ij}$: $f = \overline{F\left\{J_{ij} \right\}/N}_{J_{ij}}$.
  We accomplish this by using the replica technique
  of Edwards and Anderson \cite{Edwards_Anderson_75}.
%  and applied to the infinite ranged spin-$\frac{1}{2}$ Ising model 
%  by Sherrington and Kirkpatrick \cite{Sherrington_Kirkpatrick_75}. 
  This technique relies on
  the identity  $\ln[Z] = \lim_{n \to 0} \frac{1}{n}(Z^n - 1)$;
  $Z^n$ is interpreted as the partition function of a
  $n$-fold replicated system
  $s_i \to s_i^\alpha, \quad \alpha = 1, \ldots, n$.
  The average free energy may be computed using the prescription:

  \[\beta f = - \lim_{n \to 0} \frac{1}{n}(\overline{Z^n} - 1) . \]

  We follow this standard procedure to average the logarithm of
  the partition function. The average of the $n$-fold replicated partition
  function over the disorder gives

  \begin{eqnarray}
  \lefteqn{\overline{Z^n} =
  \tr_{ \{s^\alpha_i\} = 0,\pm 1, \dots, \pm S}
  \exp\left\{
      \frac{\beta^2 J^2}{4N} \sum_{\alpha \ne \beta}
      \sum_{ij} s_i^\alpha s_i^\beta s_j^\alpha s_j^\beta
   + (\frac{\beta^2 J^2}{4N} + \frac{\beta K}{2N})
     \sum_{\alpha}\sum_{ij}(s_i^\alpha)^2 (s_j^\alpha)^2\right.} \nonumber \\
  & & \left.
  + \frac{\beta J_0}{2N} \sum_{\alpha} \sum_{ij} s_i^\alpha s_j^\alpha
  - \beta\Delta \sum_{\alpha} \sum_{i} (s_i^\alpha)^2
  + \beta h \sum_{\alpha} \sum_{i} s_i^\alpha \right\}
  \end{eqnarray}

  The Hubbard-Stratonowich transformation
  $e^{\frac{1}{2}\lambda (\sum_\alpha s^\alpha)^2}
  = \int {\cal D} t \; e^{ t \sqrt{\lambda} \sum_\alpha s^\alpha}$
  with the convention
  ${\cal D} t = \frac{e^{-\frac{1}{2}t^2}}{\sqrt{2 \pi}} \mbox{ d} t$
  gives for the free energy, within the framework of the replica method
  at the saddle point,

  \begin{equation}
  \beta f_n = \frac{1}{2} \beta J_0 \sum_{\alpha} m_\alpha^2
      + \frac{1}{4} \beta^2 J^2 \sum_{\alpha} z_\alpha^2
      + \frac{1}{2} \beta K \sum_{\alpha} z_\alpha^2
   + \frac{1}{4} \beta^2 J^2 \sum_{\alpha \ne \beta} q_{\alpha\beta}^2
   -  \ln\left[  Z_{\rm eff} \right],
  \end{equation}

  \noindent where the effective Hamiltonian and its partition
  function is given by

  \begin{eqnarray*}
  H_{\rm eff} &=& \sum_\alpha [ - \beta \Delta  + \beta K z_\alpha
   + \frac{1}{2} \beta^2 J^2  z_\alpha ] s_\alpha^2  +
      \sum_\alpha [\beta J_0 m_\alpha + \beta h ] s_\alpha
   + \frac{1}{2} \beta^2 J^2 \sum_{\alpha \ne \beta}
        q_{\alpha\beta} s_\alpha s_\beta \\
   Z_{\rm eff} &=& \tr_{s_\alpha = 0,\pm 1,\ldots, \pm S}
   \exp\left\{H_{\rm eff}[s_\alpha]\right\}
  \end{eqnarray*}

  The quantities introduced by these transformations 
  acquire the meaning of order parameters:

  \begin{eqnarray}
   m_\alpha &=& <s_\alpha> \nonumber \\
   1 - x_\alpha = z_\alpha &=& <s_\alpha^2>  \\
   q_{\alpha\beta} &=& <s_\alpha s_\beta >, \nonumber
  \end{eqnarray}

  \noindent where the average is with respect to the effective
  Hamiltonian. These results are in fact valid for general integer
  spin values, but in what follows we restrict
  ourselves to the case of $S = 1$. In order to
  solve this model it is necessary to make assumptions on the order parameter
  matrix $q_{\alpha\beta}$ and to propose an Ansatz.

\section{Free energy of the model in RS}

%  As mentioned in the introduction, 
  We limit ourselves 
%  to extending spin variables values from
%  $s = \pm \frac{1}{2}$ to $s = 0, \pm 1$ and 
  to the simplest Ansatz for the order parameter matrix, i.e.,
  we assume symmetry with respect to permutations of any pair
  of the replicas: $q_{\alpha\beta} = q$, $\forall \alpha \ne \beta$.
  The single indexed quantities are assumed to be independent of the replica
  index:
%   as this is done even in replica breaking schemes: 
  $m_\alpha = m$ and $z_\alpha = z$, $\forall \alpha$. This leads to:

  \begin{eqnarray}\label{RSFreeEnergy}
  \lefteqn{\beta f = \frac{1}{2} \beta J_0 m^2 + \frac{1}{2} \beta K z^2
  + \frac{1}{4}\beta^2 J^2 z^2  - \frac{1}{4}  \beta^2 J^2 q^2}\\
  & & - \int {\cal D} y
   \ln\left[ 1 + 2 \cosh\left[\beta J_0 m + \beta h + \beta J \sqrt{q} y\right]
  e^{\beta K z - \beta\Delta
  + \frac{1}{2} \beta^2 J^2 z - \frac{1}{2}\beta^2 J^2 q} \right]. \nonumber
  \end{eqnarray}

  This equation reduces to the one analysed by Ghatak and Sherrington
  \cite{Ghatak_Sherrington_77}, Lage and de Almeida \cite{Lage_de_Almeida_82}
  and by Mottishaw and Sherrington \cite{Mottishaw_Sherrington_85} if
  $J_0 = 0$ and if there is no quadrupolar coupling. In order to
  simplify the notation hereafter, we define

  \begin{equation}
  \label{phi_func}
  \phi_k(y) = \frac{1}{Z_{\rm eff}(y)} \tr_{s = 0,\pm 1,\ldots, \pm S,y}
  [s^k e^{-\beta H_{\rm eff}(y)}] .
  \end{equation}

  The effective Hamiltonian and its partition function are the replica
  symmetric equivalents of those defined earlier and obtained by one
  further Hubbard-Stratonowich transformation. The set of functions $\phi_k(y)$
  reduces for the $S=1$ model to just two functions:

  \begin{eqnarray*}
  \phi_0(y) &=& \frac{2 \cosh[\beta(J_0 m + y J \sqrt{q} + h)]}{
  e^{\beta\Delta + \frac{1}{2}\beta^2 J^2 q - \beta\kappa z }
  + 2 \cosh[\beta(J_0 m + y J \sqrt{q} + h)] } \quad  \mbox{ for $k$ even,} \\
   \phi_1(y) &=& \frac{2 \sinh[\beta(J_0 m + y J \sqrt{q} + h)]}{
  e^{\beta\Delta + \frac{1}{2}\beta^2 J^2 q - \beta\kappa z }
  + 2 \cosh[\beta(J_0 m + y J \sqrt{q} + h)] } \quad  \mbox{ for $k$ odd.}
  \end{eqnarray*}

  Introducing an effective temperature-dependent quadrupolar coupling
  $\kappa = K + \frac{1}{2}\beta J^2$, the mean field saddle point
  equations in the replica symmetric approximation are:

  \begin{eqnarray}
  \frac{\partial  f}{\partial m}
  &=&  J_0 m - J_0 \int {\cal D} y \; \phi_1(y) = 0, \nonumber \\
  \frac{\partial  f}{\partial z}
  &=& \kappa z - \kappa \int {\cal D} y \; \phi_0(y) =  0, \label{z_fix_eq}\\
  \frac{\partial f}{\partial q} &=&
  - \frac{1}{2}\beta J^2  q
  + \frac{1}{2}\beta J^2
  \int {\cal D} y \left[ \phi_1(y) \right]^2 = 0. \nonumber
  \end{eqnarray}

\section{Phase diagram}

\begin{figure}[h]
  \psfig{figure=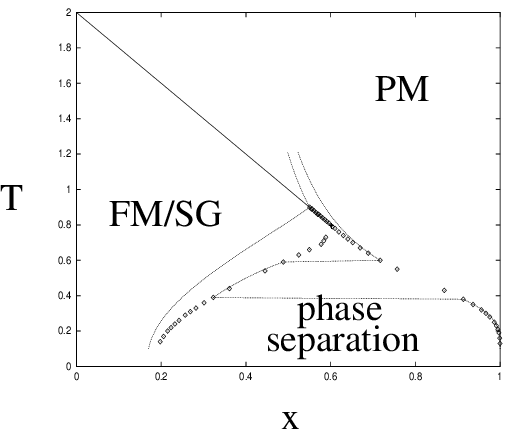,height=8cm,width=12cm}
  \caption{The phase diagram  of the disordered BEG-SK model in the
           $x-T$-plane for the bilinear couplings $J_0 = 2.0$,
           $J = 1.0$ and for weak quadrupolar coupling, $K=0.16$.
           The thin line and the dense lying diamond symbols
           represent the transition of second order, whereas
           the sparse lying diamonds indicate the first order transition.
           The two dotted lines are the free energy minima for two
           different chemical potentials ($\Delta = 0.6$ and $0.65$)
           and varying the temperature (from $T=0.0$ to $T=1.2$).}
  \label{fig:5}
  \end{figure}

  Mean field phase diagrams are obtained by extremising the free energy with
  respect to the order parameters.
  They have been determined for the BEG-SK model by
  extremising numerically the free energy (\ref{RSFreeEnergy})
  with respect to the order
  parameters $z$, $m$ and $q$ for any temperature $T$ and
  chemical potential $\Delta$. Following a line of
  constant chemical potential while varying the temperature --- as is
  shown in figure \ref{fig:5} for two different but fixed values
  of the chemical potential --- will
  occasionally reveal the onset of ordering. The examples include
  the appearence of non-vanishing
  values of the order parameters $m$ or $q$
%  (see Figure \ref{fig:5} the left of the dotted lines)
  and discontinous changes of the order parameter $z$
%  (see Figure \ref{fig:5} the right of the dotted lines), 
  which indicate a phase separation.
  We infer the transition lines 
%  of different order
  by monitoring the magnetisation $m$, the spin glass order parameter
  $q$, and the concentration $z$ for fixed chemical potential $\Delta$
  and varying the temperature $T$. This was performed for a
  range of different chemical potentials, sufficiently large to
  exhibit the different phenomenae.  

\begin{figure}[t]
  \psfig{figure=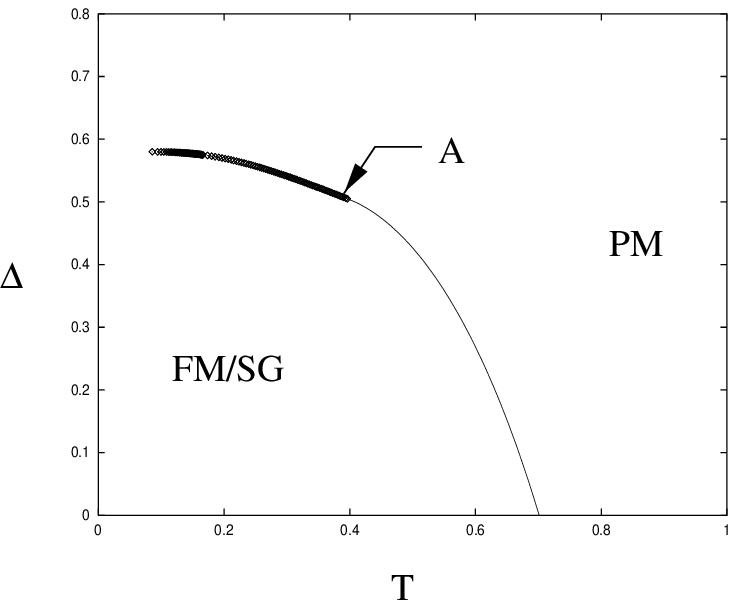,height=8cm,width=12cm}
  \caption{The phase diagram of the (non-disordered) BEG model
           in the chemical potential/temperature-plane,
           i.e., $\Delta-T$-plane, for the bilinear
           coupling $J_0 = 1.0$ and for
           weak quadrupolar coupling $K = 0.16$.
           The thin line represents a transition of second order,
           whereas the diamonds indicate a transition of first order.}
  \label{fig:2}
  \end{figure}

  To present the numerical results we will make use of two
  commonly used phase diagram sections,
  the section of chemical potential versus temperature ($\Delta$-$T$)
  and the section of concentration versus temperature ($x$-$T$).
%   \subsection{BEG model}
  For the sake of completeness we first give the results
  for the non-disordered BEG model. Since in the BEG model there
  are two order parameters, the complete phase diagram is 3-dimensional,
  but we restrict ourselves to the 2-dimensional
  sections just mentioned. In figure \ref{fig:1} we show the
  concentration/temperature ($x$-$T$) section
  and in figure \ref{fig:2} the chemical
  potential/temperature ($\Delta$-$T$) section
  of the phase diagram. Both diagrams are obtained for a small value of the
  quadrupolar coupling constant $K=0.16$. 

\begin{figure}[t]
  \psfig{figure=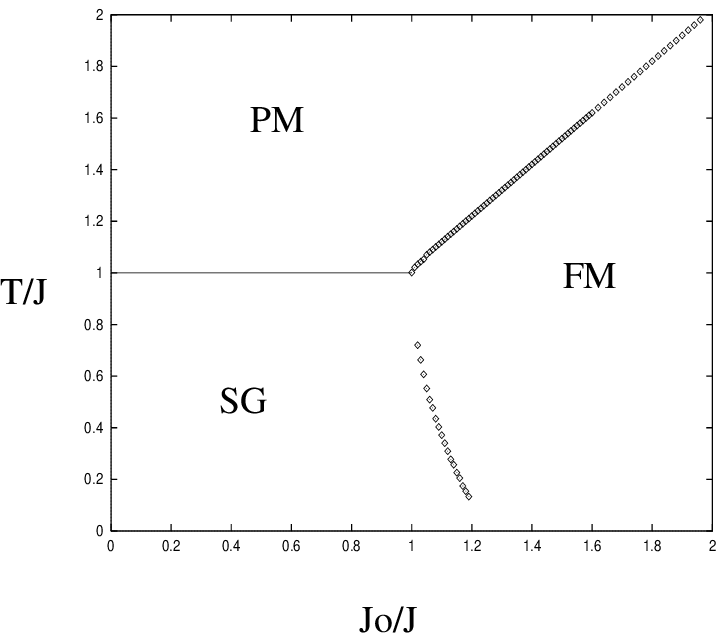,height=8cm,width=12cm}
  \caption{The phase diagram of the SK model in the ferromagnetic
           coupling/temperature-plane, i.e., $J_0/J-T$-plane.
           The lines shown are transitions of second order.}
  \label{fig:3}
  \end{figure}

%  \subsection{SK model}
  Next, we consider a special limit of the BEG-SK model to
  make contact to the work of Sherrington and Kirkpatrick.
  We allow the
  concentration to vary freely and we take the limit $\Delta \to -\infty$.
  From this, we recover the Sherrington
  and Kirkpatrick Hamiltonian \cite{Sherrington_Kirkpatrick_75}: all
  spins tend to be $|s| = 1$, i.e., $x \to 0$. The phase diagram obtained
  by Sherrington and Kirkpatrick represents
  a section of the general phase diagram, i.e., the $x$-$J_0$-$T$ section
  for a concentration value $x=0$.
  This diagram exhibits a {\em paramagnetic} phase at high temperatures
  and small ferromagnetic coupling, whereas it exhibits a
  {\em ferromagnetic} phase
  for larger ferromagnetic coupling and moderate temperatures.
  Finally at very low temperatures the spin glass phase appears.
  This behaviour is shown in figure \ref{fig:3}. The figure does
  not take into account the correction to the transition line between
  ferromagnetic and spin glass phase due to stability requirements
  \cite{de_Almeida_Thouless_78}.

%\subsection{GS-model}
  The section of the general phase diagram obtained by Ghatak and
  Sherrington \cite{Ghatak_Sherrington_77}
  is the $\Delta-T$-section for $J_0=0$, cf. \cite{Abdelouahab_94}.
  The transition lines are obtained analytically following
  the procedure of Lage
  and de Almeida \cite{Lage_de_Almeida_82} by examining the stability
  of the replica symmetric solution; they are given by

  \begin{equation}
  \Delta = \left\{ \begin{array}{r@{\quad:\quad}l}
        \frac{J}{2} +
        T \ln\left[ 2 \left(\frac{J}{T}-1\right)\right]
        & T > \frac{1}{3} \\
        \frac{J^2}{4 T}\left(1 \pm \sqrt{1 - \frac{8T^2}{J^2}}\right) +
        T \ln\left[2 \left( \frac{1}{\frac{1}{2}
        (1 \pm \sqrt{1 - \frac{8T^2}{J^2}})} - 1\right)\right]
        & T \le \frac{1}{3} \\
        \end{array}\right. .
  \end{equation}

\begin{figure}[h]
  \psfig{figure=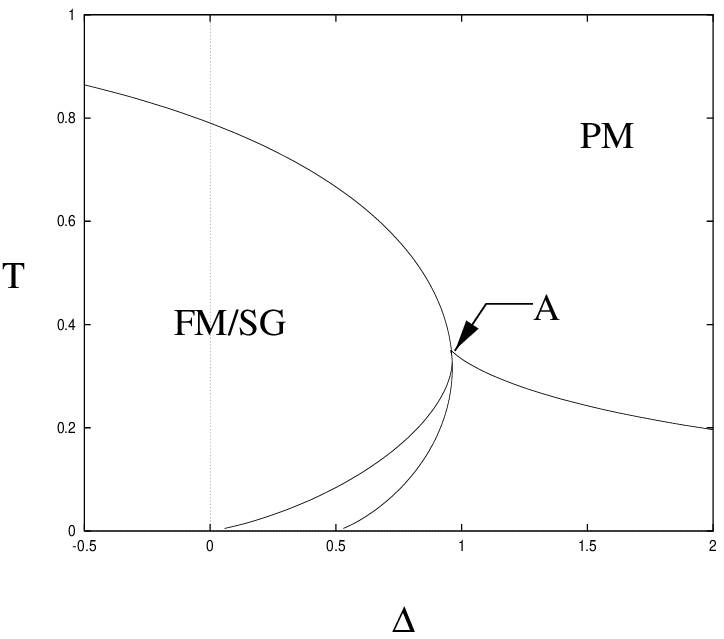,height=8cm,width=12cm}
  \caption{The phase boundary of the GS-model ($K = 0.0$ and
           $J_0 = 0.0$) in the chemical potential/temperature-plane,
           i.e., $\Delta-T$-plane.}
  \label{fig:4}
  \end{figure}

  \noindent These transition lines are shown in figure \ref{fig:4}.
  The lower two lines belong respectively to the two choices of sign
  in the second equation.

\begin{figure}[h]
  \psfig{figure=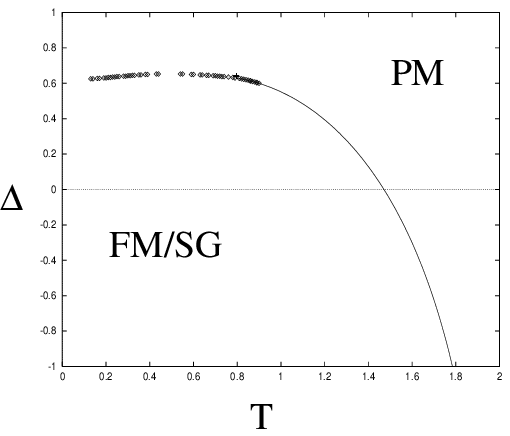,height=8cm,width=12cm}
  \caption{The phase diagram  of the disordered BEG-SK model
           in the $\Delta$-$T$-section for the bilinear
           couplings $J_0 = 2.0$, $J = 1.0$ and for weak
           quadrupolar coupling $K=0.16$.
           The thin line and the dense lying diamond symbols
           represent the second order transition,
           whereas the sparse diamonds indicate the transition of first order.}
  \label{fig:6}
  \end{figure}

\begin{figure}[t]
  \psfig{figure=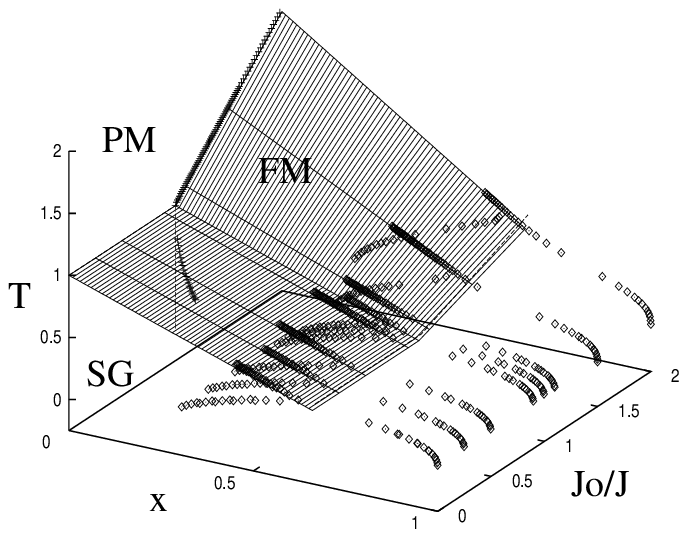,height=8cm,width=12cm}
  \caption{The complete phase diagram of the disordered BEG-SK model,
           with the lines of stability and the line of tricritical
           temperature in the concentration/ferromagnetic
           coupling/temperature-representation, i.e.,
           $x-J_0/J-T$-plane, for weak quadrupolar coupling
           $K = 0.16$.}
  \label{fig:7}
  \end{figure}

\begin{figure}[t]
  \psfig{figure=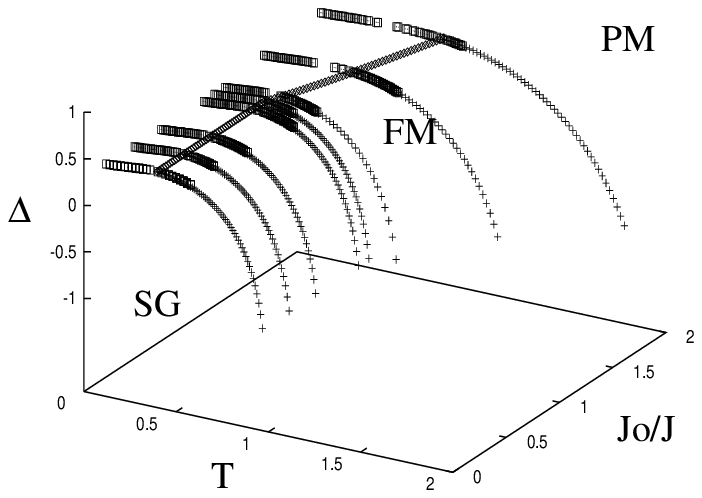,height=8cm,width=12cm}
  \caption{The complete phase diagram of the disordered BEG-SK model,
           with the lines of stability and the line of tricritical
           chemical potential in the  chemical potential/ferromagnetic
           coupling/temperature-representation, i.e.,
           $\Delta-J_0/J-T$-plane, for weak quadrupolar coupling $K = 0.16$.}
  \label{fig:8}
  \end{figure}

\begin{figure}[h]
  \psfig{figure=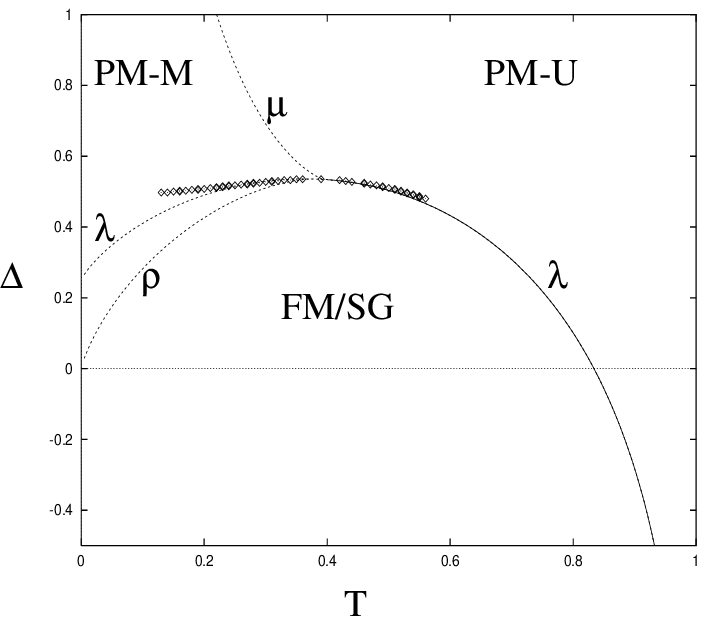,height=8cm,width=12cm}
  \caption{The phase diagram of the disordered BEG-SK model,
       with the lines of stability in
       the chemical potential/temperature-representation, i.e.,
       $\Delta-T$-plane, for weak quadrupolar coupling,
       $K = 0.16$, and for the ferromagnetic coupling ${J_{0}}/{J}=2.0$.
       The symbols have the following meanings: PM-M is the paramagnetic
       phase with many free energy minima, PM-U is the paramagnetic
       phase with an unique free energy minimum, FM/SG the ferromagnetic
       or eventually spin glas phase, $\mu$ gives the positive sign branch
       and $\varrho$ the negative sign branch of equation (\ref{DeltaRS});
       $\lambda$ is the second order transition line.}
  \label{fig:11}
  \end{figure}

%\subsection{Complete phase diagram}
  Finally we are ready to present 
% give the results of this work, i.e., 
  the complete phase diagram for BEG-SK model.
  Figure \ref{fig:5} shows the
  $x$-$T$ section and figure \ref{fig:6} the $\Delta$-$T$ section
  of the phase diagram for one choice of the coupling
  parameters: $J_0 = 2.0$, $J = 1.0$ and $K = 0.16$.
  In order to relate this to previous work and to present the
  complete phase diagram, the $x$-$T$ and the $\Delta$-$T$ sections have been
  calculated for different values of the ferromagnetic coupling (scaled
  by the variance of the distribution of the couplings) and the results
  are shown in figures \ref{fig:7} and \ref{fig:8}, respectively.
  Figure \ref{fig:7} shows in the $x$-$J_{0}/J$-$T$ plane our
  numerical results (represented by diamonds) and the results
  of the SK model (represented by crosses) at a value $x=0$ of 
  the concentration.
  The line of tricritical temperatures which will be given in
  formula (\ref{tricrittemp})
  is shown with a broken line and the surface gives the second order
  phase transition. In figure \ref{fig:8} the $\Delta$-$J_{0}/J$-$T$
  section is given and the square symbols indicate the
  numerical results, which cover the first as well as the second order
  phase transition. The results from Ghatak and Sherrington
  \cite{Ghatak_Sherrington_77} are reproduced for the value of the
  ferromagnetic coupling $J_{0}= 0$, i.e., the face of the cube.
  For details on the stability we
  refer the reader to section \ref{stabilitydiscussion} and to figure
  \ref{fig:11}. The first order transitions
  terminate at the line of tricritical potential given by formula
  (\ref{critdelta}), which is
  the thick line from the bottom left to the top right. The crosses
  represent the first order transition lines. The second order
  transition above $T_{\rm tri}$,
  given by formula (\ref{secondorderstabilityanalysis})
  and obtained by the stability analysis
  presented in Section 6 are represented by squares.
  The stability lines below $T_{\rm tri}$ are not given.

  We now discuss the numerical results in the light of recent
  work on disordered BEG-type models which are not frustrated.
  After we will address the problem of the stability of the replica
  symmetric solution.

\section{Behavior of the first order transition}

  Lets look at the influence of disorder on the
  first order transition, which can --- following Blume, Emery and
  Griffiths --- be interpreted as a phase separation. This first
  order transition represents
  an example of a transition which breaks a global symmetry. The
  Hamiltonian is invariant under inversion of all spins;
  in the ordered phase this symmetry is
  spontaneously broken, whereas in the paramagnetic phase the
  symmetry is preserved.

\begin{figure}[t]
  \psfig{figure=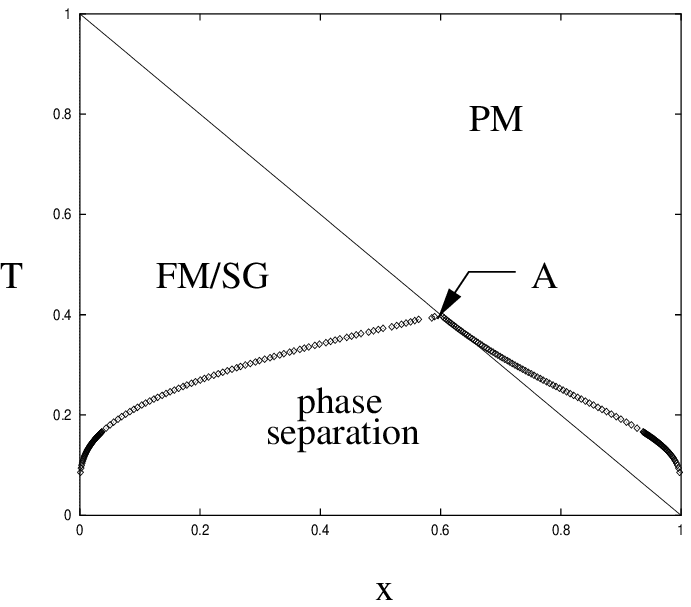,height=8cm,width=12cm}
  \caption{The phase diagram of the (non-disordered) BEG model in the
           concentration/temperature-plane, i.e., $x-T$-plane,
           for the bilinear coupling $J_0 = 1.0$ and for
           weak quadrupolar coupling $K = 0.16$.
           The thin line represents a transition of second order,
           whereas the diamonds indicate a transition of first order.
           FM/SG stands for ferromagnetic/spin glass phase,,
           PM is the paramagnetic phase;
           the point A indicates the tricritical point.}
  \label{fig:1}
  \end{figure}

\begin{figure}[h]
  \psfig{figure=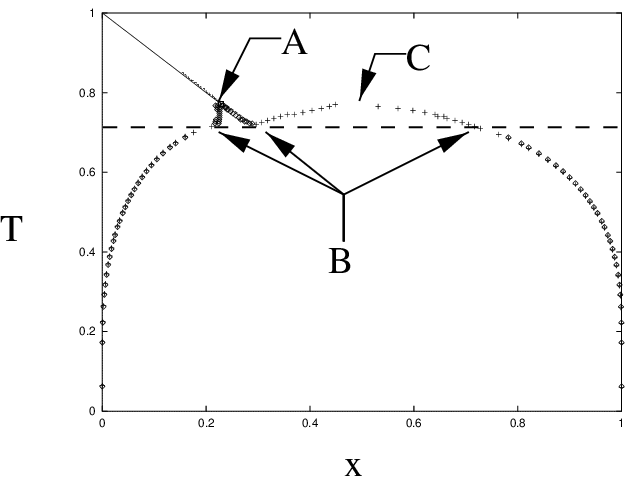,height=8cm,width=12cm}
  \caption{The phase diagram of the non-disordered BEG model in the
            $x-T$-plane for the bilinear
            coupling $J = 1.0$ and for a rather strong
            quadrupolar coupling, $K = 2.88$.
            The point A indicates the tricritical point,
            the point B represents the triple point and
            point C is the critical end point.}
  \label{fig:9}
  \end{figure}

\begin{figure}[h]
  \psfig{figure=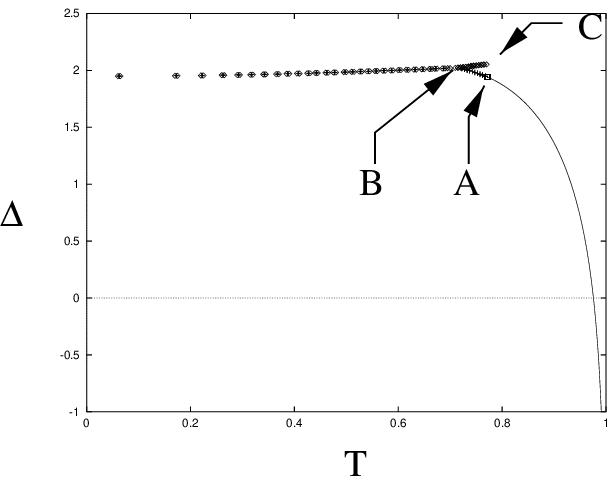,height=8cm,width=12cm}
  \caption{The phase diagram of the non-disordered BEG model in the
            $\Delta-T$-section for the bilinear
            coupling $J = 1.0$ and for a rather strong
            quadrupolar coupling $K = 2.88$.}
  \label{fig:10}
  \end{figure}

  We first consider the non-disordered BEG model in
  the $x-T$ representation, where the appearance of the phase
  separation is most visible. In figure \ref{fig:1}
  the phase diagram is shown for a weak quadrupolar interaction.
  In figure \ref{fig:9} the same is seen for a rather strong
  quadrupolar coupling $K = 2.88$. The point is that the
  quadrupolar coupling enhances the phase separation.
  In the limit of $K >> 1$, the model tends to behave as the
  Griffiths model \cite{Griffiths_67}. We thus wish to emphasise
  the importance of the presence of a quadrupolar interaction.
  In the original model of Blume, Emery, and Griffiths, even the $K = 0$ case
  exhibits a phase separation. It is due to the bulk interaction of the
  $s=0$ species (He$^3$) and the $s=\pm 1$ species (He$^4$), the latter
  allowing an additional degree of freedom (ferromagnetic phase).
  At low temperature the ferromagnetic phase is the favoured phase
  for the $s=\pm 1$ species and phase separation occurs in order
  to permit ordering in a He$^4$ rich phase. Letting $K \ne 0$
  allows for an inter-isotopic interaction; in fact $K$
  represents a combination of inter-isotopic interactions,
  $K = K_{33} + K_{44} - 2 K_{34}$, which are assumed
  to be positive. 
%   This interaction is combined with the bulk interaction. 
  It is remarkable that for a range of values of $K$
  the phase diagram exhibits a triple point together with the critical
  and tricritical point. This appears in the original BEG model
%  confer figure \ref{fig:10} for the appearence of a triple point, marked as A,
  and in the present model, as can be seen from figure \ref{fig:10}.
  The tricritical point is given in the diagrams by {\em A}, the triple
  point by {\em B}, and the critical point by {\em C}; these special points
  are also marked in figure \ref{fig:9}.

  Introducing disorder in the BEG model may affect the
  phase separation. Figure \ref{fig:5} shows the phase diagram
  obtained by a numerical extremisation of the free energy.
  {\em The phase separation persists independently of the
  presence of disorder}. The average over the Gaussian
  disorder has introduced an effective quadrupolar coupling
  $\kappa$, which --- as we have seen before in the non-disordered
  case ---  enhances the phase separation. Furthermore this
  effective interaction is also temperature dependent:

  \begin{equation}
  \kappa = K + \frac{1}{2}\beta J^2.
  \end{equation}

  This promotes the phase separation
  at low temperatures as can be seen by comparing
  with figure \ref{fig:9} (for the low temperature
  regime) and figure \ref{fig:1} (for the high temperature regime)
  in the non-disordered case. Even if there is an effect
  due to the presence of disorder, which might suppress or
  change the first order transition,
  this effect is apparently compensated by the additional
  term contributing to the effective quadrupolar coupling.

  This result complements those of Berker {\em et al.}
  \cite{Berker_91}, \cite{Berker_93} and \cite{Berker_Falicov_95_b}.
  They conjecture, in dimension $d \ge 3$, that the disorder
  lowers the tricritical temperature; then that part of the
  transition line enclosed between the former and the actual
  tricritical temperature should become second order.
  Also, they claim that all of the first order line is
  replaced by a second order
  one if the disorder is sufficiently strong.
  A few comments are in order:
  \begin{enumerate}
  \item[({\em i}\/)] The analysis of Berker {\em et al.}
  relies on the real space
  renormalisation group approach for an initially positive
  distribution of the couplings. This distribution remains
  positive and consequently cannot take into account the effect
  of frustration. Nevertheless, we do not think that the frustration
  is responsible for the persistence of the phase separation
  for the following reason: even when $J$ is small, so that there
  is disorder but essentially no frustration, the phase separation
  persists (see the phase diagrams). Monte Carlo studies
  by Diep {\it et al.} \cite{Diep_97} indicate this persistence and hence
  ensure that the
  phase separation is not an artefact of the mean field approach.
  \item[({\em ii}\/)] The mean field approximation is equivalent
  to a model in infinite
  dimensions, perhaps rendering the comparison of our results with
  the predictions of Berker {\em et al.} invalid. Nevertheless, 
  it should give an indication of whether the first order transition
  persists or not. Furthermore one could argue that in infinite
  dimensions, an infinitely strong disorder is required to suppress
  the first order transition. However figure \ref{fig:4}
  indicates that this model reduces to that discussed by Ghatak
  and Sherrington \cite{Ghatak_Sherrington_77} with a vanishing
  ferromagnetic coupling, showing again the persistence of the
  first order transition.
  \item[({\em iii}\/)] The assumption of replica symmetry is not justified.
  However, based on the stability analysis, we have found that the first
  order transition persists; it seems rather improbable
  that it be suppressed by doing the full replica symmetry breaking
  scheme. 
%  In order to clarify this point and others concerning the
%  validity of the replica Ansatz, work is progress
%  to perform the complete replica symmetry breaking scheme.
  \end{enumerate}

  To summarise, Berker {\em et al.} claim that their results are
  generic, but
  our mean field treatment shows that the phase separation is not
  affected by the introduction of disorder. In particular, the tricritical
  temperature is not modified: ${T}/{J} = {1}/{3}$, if $J> J_0$,
  using the same scaling as Berker {\em et al}.
  Note that their analysis does not reveal
  the effective quadrupolar interaction, which plays a subtle role.
  On the one hand, following Berker {\em et al.} and their general arguments
  \cite{Berker_91,Berker_93}, the disorder should change
  the first order transition to a second order one; on the other hand,
  the effective quadrupolar interaction enhances the phase separation,
  so stabilises the first order transition. As we have seen above, the
  effective quadrupolar interaction more than compensates the first effect
  so that the first order transition persists. Furthermore,
  since for a small amount of disorder the frustration effects 
  are negligible, the models considered are comparable and our results
  provide a counterexample to their claim.

\section{Transition lines from a stability analysis}

%  The application of the replica method to the
%  problem of random, but quenched interactions \cite{Edwards_Anderson_75}
%  was a first step in understanding spin glass magnetic ordering.
%  The replica symmetric solution of the infinite-range SK-Ising model due to
%  Sherrington and Kirkpatrick
%  \cite{Sherrington_Kirkpatrick_75} suffers from several drawbacks,
%  such as for example negative entropy in the low temperature region.
%  Later de Almeida and Thouless \cite{de_Almeida_Thouless_78} showed
%  that the solution obtained by assuming a symmetric Ansatz may be unstable,
%  implying the need to break the replica permutation symmetry. The currently
%  accepted replica symmetry breaking scheme, developed by
%  Parisi \cite{Parisi_80}, has been shown to remove the instabilities
%  in all cases studied to date.
  The assumption of replica symmetry allows for a simple answer
  for the BEG-SK model. However, as for the
  SK model, it
%  by de Almeida and Thouless \cite{de_Almeida_Thouless_78}, 
%  the assumption of replica symmetry 
  leads to an unstable solution and the breaking 
  of the permutation symmetry of
  the replicas is required. 
%  The same is true for the BEG-SK model and
  We expect the Parisi breaking scheme 
%  may equally well be applied here to remove
  to apply here and to remove the instabilities in our model.
  Since under replica symmetry breaking the findings of the
  analysis assuming replica symmetry are in general confirmed,
  {\it i.e.}, a spin glass phase persists, we won't study finite
  nor infinite step step replica symmetry breaking.
  In this section we restrict ourselves to the replica symmetric solution
  and we derive the lines of instability to locate the phase transitions lines
  given in section IV
  --- even though the location of these lines may differ in a more advanced
  treatment.
  The stability of the replica symmetric solution of the SK model in its
  integer spin generalisation was examined
  by Lage and de Almeida \cite{Lage_de_Almeida_82} and in greater detail
  by Mottishaw and Sherrington \cite{Mottishaw_Sherrington_85}.
  These works limit their analysis to the disordered integer S spin
  glass model without a non-zero ferromagnetic mean and without
  quadrupolar coupling.

  The present stability analysis follows closely the moethods
  of de Almeida and
  Thouless \cite{de_Almeida_Thouless_78} and subsequent works. To examine
  the stability of the solution it has to be made sure that the solution
  extremises the free energy. The quadratic form $\Gamma$ describing the
  deviation of the solution from its stationary value should be
  positive definite.

  \begin{eqnarray}
  \beta f &=&  \beta f(m,z,q) - \frac{1}{2} \Gamma + {\cal O}(\delta^3)
  \quad\mbox{and} \nonumber \\
  \Gamma &=&
  \sum_{\alpha\beta} G_{\alpha\beta}^{\epsilon\epsilon}
        \epsilon_\alpha \epsilon_\beta +
  \sum_{\alpha\beta} G_{\alpha\beta}^{\rho\rho}
        \rho_\alpha \rho_\beta +
  \sum_{(\alpha\beta)(\gamma\delta)} G_{(\alpha\beta)(\gamma\delta)}^{\eta\eta}
     \eta_{\alpha\beta} \eta_{\gamma\delta} \\
  &+& 2 \sum_{\alpha\beta} G_{\alpha\beta}^{\epsilon\rho}
        \epsilon_\alpha \rho_\beta +
  2 \sum_{(\alpha\beta) \gamma} G_{(\alpha\beta)\gamma}^{\eta\epsilon}
        \eta_{\alpha\beta} \epsilon_\gamma +
  2 \sum_{(\alpha\beta) \gamma} G_{(\alpha\beta)\gamma}^{\eta\rho}
        \eta_{\alpha\beta} \rho_\gamma.    \nonumber
  \end{eqnarray}

  \noindent
  The matrix $G$ associated with this form is the Hessian ${\cal H}$. The
  eigenvalues of the Hessian should be non-negative to make sure that
  the solution is stable. Due to the same symmetry arguments used by
  de Almeida and Thouless \cite{de_Almeida_Thouless_78},
  we discover three families of eigenvalues which by convention are called:
  {\em longitudinal}, {\em longitudinal anomalous} and {\em replicon}.
  But in the present case the stability analysis requires the
  diagonalisation of a Hessian ${\cal H}$ which is built out of
  three blocks. As a consequence there are now three
  distinct eigenvalues in each family.
  The longitudinal eigenvectors are the fully symmetric ones with respect to
  permutations of the replica indices and are of the form:

  \begin{equation}
  e_\mu^L = a \mbox{ for } \mu = 1\ldots n,
  e_\nu^L = b \mbox{ for } \nu = 1\ldots n,
  e_{\alpha\beta}^L = c \mbox{ for }\alpha,\beta = 1\ldots \frac{n}{2}(n-1).
  \end{equation}

  They give rise to eigenvalues which are the solution to the cubic
  characteristic equation. The full formula for the eigenvalues is too
  cumbersome to be displayed here, they will be discussed in the paramagnetic
  phase below (the full expressions are given in appendix
  \ref{eigenapp}).
  The longitudinal anomalous
  eigenvectors are generated by one distinct replica index $\theta$ and
  look like:

  \begin{eqnarray}
  e_\mu^{LA} &=& a \mbox{ for } \mu = 1\ldots n
  \mbox{ and } \mu \ne \theta,
  e_\mu^{LA} = g \mbox{ for } \mu = \theta, \nonumber \\
  e_\nu^{LA} &=& b \mbox{ for } \nu = 1\ldots n
  \mbox{ and } \mu \ne \theta,
  e_\nu^{LA} = c \mbox{ for } \nu = \theta,\\
  e_{\alpha\beta}^{LA} &=& d
  \mbox{ for } \alpha,\beta = 1\ldots \frac{n}{2}(n-1)
  \mbox{ and } \alpha,\beta \ne \theta,
  e_{\alpha\beta}^{LA} = e \mbox{ for } \alpha,\beta = \theta. \nonumber
  \end{eqnarray}

  Again the eigenvalues are given in appendix \ref{eigenapp}. The
  replicon eigenvectors
  are generated by two distinct replica indices $\omega$ and $\theta$
  and they are of the following form:

  \parbox{11cm}{\begin{eqnarray*}
  e_\mu^R &=& a \mbox{ for } \mu = 1\ldots n
  \mbox{ and } \mu \ne \theta,\omega,
  e_\mu^R = g \mbox{ for } \mu = \theta,\omega,\\
  e_\nu^R &=& b \mbox{ for } \nu = 1\ldots n
  \mbox{ and } \mu \ne \theta, \omega,
  e_\nu^R = c \mbox{ for } \nu = \theta, \omega,\\
  e_{\alpha\beta}^R &=& d \mbox{ for }
  \alpha, \beta = 1\ldots \frac{n}{2}(n-1)
  \mbox{ and } \alpha,\beta \ne \theta
  \mbox{ and } \alpha,\beta \ne \omega,\\
  e_{\alpha\beta}^R &=& e
  \mbox{ for } \alpha, \beta = \theta \mbox{ or }
  \alpha, \beta = \omega,
  e_{\alpha\beta}^R = f \mbox{ for } (\alpha \beta) = (\theta\omega).
  \end{eqnarray*}} \hfill
  \parbox{1cm}{\begin{eqnarray}\end{eqnarray}}

  The eigenvalue is given by the solution of the characteristic equation,
  but as this equation is explicit, so the eigenvalue is obtained immediately:

  \begin{equation}
  \lambda_R = P - 2 Q + R,
  \end{equation}

  \noindent
  where $P, Q$ and $R$ are appropriate correlation functions given in
  the appendix.
  Adding the number of eigenvalues obtained in the three symmetry
  families gives the expected number
  $3 + (3n - 3) + \frac{n}{2}(n-3) = \frac{n}{2}(n+3)$,
  i.e., the dimension of ${\cal H}$. The so called longitudinal eigenvalues
  $\lambda_L$ and $\lambda_{LA}$ coincide in the $n \to 0$-limit
  as in the work of de Almeida and Thouless
  \cite{de_Almeida_Thouless_78} and Lage and de Almeida
  \cite{Lage_de_Almeida_82}. The replica symmetric fluctuations of
  the one-indexed quantities ($z_\alpha$ and $m_\alpha$) and the
  two-indexed quantity $q_{\alpha\beta}$ are described respectively
  by the longitudinal eigenvalues, $\lambda_{LA}$ and $\lambda_L$.
  The third eigenvalue $\lambda_R$ is distinct
  from the other two eigenvalues and is related to the
  fluctuations which break the replica symmetry of the two indexed
  quantity $q_{\alpha\beta}$. In order to investigate the stability of
  the replica symmetric solution, for example in the ferromagnetic phase,
  the eigenvalues $\lambda_L$, $\lambda_{LA}$ and $\lambda_R$ have to be
  calculated in this phase. In the appendix the general expressions
  for the different eigenvalues and the correlation
  functions appearing therein are given. They must be evaluated
  under the appropriate approximation, i.e., in the present case,
  under the replica symmetric approximation. In order for a phase to be
  stable the eigenvalues in this phase have to be non-negative.
  We now investigate the stability of the paramagnetic phase.

\subsection*{The paramagnetic phase}

  Consider the paramagnetic phase defined by the vanishing 
  of all magnetic order
  parameters: $q = 0, m = 0$.
  This considerably simplifies the expressions for the eigenvalues.
  All off-diagonal elements of the Hessian or combinations appearing
  in the characteristic equations vanish.
  In order to simplify the notation further, we consider directly the
  combinations appearing in the definitions of the different eigenvalues.
  The {\em longitudinal} eigenvalues are given by the solution of:

  \begin{equation}
  \lambda_L^3 + a \lambda_L^2 + b\lambda_L + c
  = (\lambda-\lambda_1)(\lambda-\lambda_2)(\lambda-\lambda_3) = 0.
  \end{equation}

  The coefficients for a cubic equation are given implicitly by the
  solutions to that equation:

  \begin{eqnarray}
  - a &=& [U-V] + [A-B] + [P-4\,Q+3\,R]
       = \lambda_1 + \lambda_2 + \lambda_3 \nonumber \\
  b &=& [U-V] [P-4\,Q+3\,R] + [A-B] [P-4\,Q+3\,R] + [U-V] [A-B]
    = \lambda_1\lambda_2+\lambda_2\lambda_3 + \lambda_1\lambda_3 \\
  c &=& -[U-V] [A-B] [P-4\,Q+3\,R] = -\lambda_1 \lambda_2 \lambda_3, \nonumber
  \end{eqnarray}

  \noindent
  which may be solved immediately to give

  \begin{eqnarray}
  \lambda_1^L &=& [A-B] \nonumber \\
  \lambda_2^L &=& [U-V] \\
  \lambda_3^L &=& [P-4\,Q+3\,R] = P. \nonumber
  \end{eqnarray}

  The {\em longitudinal anomalous} eigenvalues are given by the same
  expressions. The {\em replicon} eigenvalue is given by:

  \begin{equation}
  \lambda_R = P - 2 Q + R = P,
  \end{equation}

  \noindent
  and coincides in the paramagnetic phase with the third longitudinal
  eigenvalue. In order for the paramagnetic phase to be stable,
  the eigenvalues in this phase must be non-negative.
  The border line of stability is given by $\lambda_{L, LA, R} = 0$,
  which defines seven lines; but due to the collapsing eigenvalues of
  different families there are only three different stability lines.
  In the replica symmetric approximation, these read:

  \begin{eqnarray}
  A-B &=& \beta J_0 [1 - \beta J_0 z] = 0 \nonumber \\
  U-V &=& \beta \kappa [1 - \beta \kappa (z - z^2)] = 0\\
  P-4\,Q+3\,R = P - 2 Q + R &=& \beta^2 J^2  [ 1 - \beta^2 J^2 z^2 ]
   = 0 \nonumber .
  \end{eqnarray}

  The stability limits are given by the following explicit formulae,
  first in the plane of concentration $x$ versus temperature $T$:

  \begin{eqnarray}
  \label{secondorderstabilityanalysis}
  \frac{T(x)}{J_0} &=& 1 - x \nonumber \\
  \frac{T(x)}{J} &=& 1 - x \\
  \frac{1}{\beta K + \frac{1}{2} \beta^2 J^2} &=& x (1-x) \nonumber .
  \end{eqnarray}

  \label{stabilitydiscussion}
  The paramagnetic phase becomes unstable whenever the temperature is
  lowered below the greater of the two bilinear couplings $J$ and $J_0$.
  This behaviour is identical to that observed by Sherrington and Kirkpatrick.
  The conditions of stability can be rewritten in terms of the
  chemical potential $\Delta$ and temperature $T$. The stability lines
  are obtained by using the fixed point equation (\ref{z_fix_eq})
  determinating the concentration, and represent a line in the
  $\Delta$-$T$-section of the phase diagram.

  \begin{equation}\label{DeltaRS}
  \Delta_{\rm RS}(T) = \left\{ \begin{array}{r@{\quad:\quad}l}
        (K + \frac{1}{2}\beta J^2) \frac{T}{J} +
        T \ln\left[ 2 \left(\frac{J}{T}-1\right)\right] & T > T_{\rm tri} \\
        (K + \frac{1}{2}\beta J^2) \frac{1}{2}
        \left(1 \pm \sqrt{1 - \frac{8}{2\beta K + \beta^2 J^2}}\right) +
        T\ln\left[2 \left(\frac{1}{\frac{1}{2}
        (1 \pm \sqrt{1 - \frac{8}{2\beta K + \beta^2 J^2}})} - 1\right)\right]
        & T \le T_{\rm tri}  \\
        \end{array}\right. .
  \end{equation}

  For one choice of the
  parameter the complete set of stability lines has been depicted
  in figure \ref{fig:11}. Furthermore the numerical results covering the
  first order transition points as well as those of second order are
  represented by diamond symbols. The following discussion refers to
  this figure.
  The two choices of sign in the second equation separate different
  regions of the free energy's behaviour.
  The branch ($\mu$) belonging to the positive sign separates the region of
  free energy with a unique minimum (PM-U) from the region where there is more
  than one local minima (PM-M). It is in this latter region where first order
  transitions occur and the phase separation appears.
  The branch ($\varrho$) belonging to the negative sign represents
  regions where the free energy has three local minima, separating
  the region where the minimal solution is the one with vanishing
  order parameter from the region where the minimal solution is
  the one with a nonvanishing order parameter; this line ($\varrho$)
  is indicating
  the onset of phase separation. The concentration $x$ jumps when crossing
  this line; in fact this line represents in the $\Delta$-$T$ section of
  the phase diagram the whole phase separation or coexistence region. As
  can be seen by comparision with the numerical data, the location of the
  first order phase transition is predicted correctly only near the
  tricritical point. It is known that the stability analysis is not an adequate
  mean to determine first order phase transitions. As in previous
  works the $\lambda$-line ($\lambda$) meets the first order transition
  line at a tricritical point: a second order line changes to a first
  order line, or, following Griffiths
  \cite{Griffiths_70}, three critical lines meet (see also
  \cite{Lawrie_Sarbach_83}). For completeness the continuation of
  the $\lambda$-line below $T_{\rm tri}$ has been drawn too.

\subsection*{Tricriticality}

  The occurrence of a tricritical point in a system with one order
  parameter is signaled by the vanishing of the fourth derivative
  of the free energy with respect to the order parameter. In the
  present problem there are three order parameters and they are
  have replica indices. The criterion must be modified:
  the eigenvalues of the matrix of the fourth derivatives have to
  vanish. But due to the diagonal structure of the Hessian ${\cal H}$
  in the paramagnetic phase this amounts to 
  computing the fourth derivative with respect to the magnetisation
  and to search for its zero. This is in fact the classical argument
  of the Landau theory.
  The tricritical temperature in the non-disordered case is given for
  a vanishing quadrupolar coupling by
  
  \[ T_{\mbox{\rm\scriptsize tri}} =
  \frac{1}{3}\max \left\{J_{0}, J\right\}. \]

  For a non-vanishing quadrupolar coupling the tricritical temperature
  is given by:

  \[  T_{\mbox{\rm\scriptsize tri}} =
  \frac{2K + 1}{2K + 3}\max \left\{J_o, J\right\}. \]

  The tricritical point present in the non-disordered model (BEG) persists
  and in terms of the temperature and concentration is given in the crudest
  approximation by:

  \begin{equation}
  \label{tricrittemp}
  x_{\mbox{\rm\scriptsize tri}} = \frac{2}{3} \quad \mbox{and} \quad
  T_{\mbox{\rm\scriptsize tri}} = \frac{2K + 1}{2K + 3}
  \max \left\{ J_{0}, J \right\} .
  \end{equation}

  The numerical data show that this is a 
  good approximation. It is worth
  mentioning that the tricritical concentration has kept its
  value from the non-disordered BEG model. In contrast the tricritical
  chemical potential has been modified. Resolving
  the tricritical condition provides a formula in
  $\Delta$-$J_0/J$-$T$ space for the tricritical line.
  In the $\Delta$-$T$ plane, it is given by:

  \begin{equation}
  \label{critdelta}
  \Delta_{\rm tri}(J_0, J) = \left\{ \begin{array}{rl}
  \frac{K}{3} + \frac{1}{2} \frac{J^2}{J_0} + \frac{2}{3} J_0 \ln[2]
  & \mbox{ and } \quad
  T_{\rm tri} = \frac{J_0}{3} \mbox{ for } J_0 > J\\
  \frac{K}{3} + \frac{1}{2} J + \frac{2}{3} J \ln[2]
  & \mbox{ and } \quad
  T_{\rm tri} = \frac{J}{3}  \mbox{ for } J > J_0 \\
  \end{array}\right. .
  \end{equation}

  This line, as a function of $J_0$, is constant up to $J_0 = J$, and then
  tends for stronger ferromagnetic coupling to larger chemical potential.
  The line of stability for the second order transition and the
  aforementioned line of tricritical $\Delta$ are displayed
  together with the numerical data in figure \ref{fig:8}. The value
  for the quadrupolar coupling for this figure is $K=0.16$.

\section{Discussion and Conclusions}

  We have performed a replica study of the disordered BEG model and
  have extended previous work to present a picture of the complete
  phase diagram. The BEG-SK model shows a second order transition,
  the so-called $\lambda$-line, separating an ordered phase from a
  paramagnetic one. The ordered phase
  may be a ferromagnetic or a spin glass phase, depending on the strength
  of the ferromagnetic coupling $J_0$. Furthermore, a transition of first
  order, present in the non-disordered BEG model, persists, and may
  be interpreted as a phase separation. This extends recent work of
  Berker {\em et al.} on the influence of disorder on first order
  transitions. Our mean field study is completed by a stability
  analysis of the replica symmetric approximation. The
  complete set of eigenvectors and eigenvalues has been found and
  analysed in the paramagnetic phase. The replica symmetric solution
  suffers from instabilities, but the assumption of replica symmetry
  and the associated results do {\em not} exhibit more severe
  inconsistencies than in the SK model. For the SK model, the
  instabilities of the replica symmetric solution have been remedied
  by the infinite step replica symmetry breaking procedure. We
  expect the same approach to work in this case, but following Mottishaw and
  Sherrington \cite{Mottishaw_Sherrington_85} near the tricritical
  point, the Parisi Ansatz has to be extended to higher order in
  $q_{\alpha\beta}$.

\section{Acknowledgement}

  I would like to thank Thomas Garel for suggesting this
  problem, T. H. Diep, C. de Dominicis,
  O. Martin, H. Orland for helpful discussions, and
  A. Sedeki for checking some of the computatations.
  An Individual Fellowship of the Commission of European Communities
  under contract number ERBCHBICT941665 is gratefully acknowledged.
  Furthermore, I would like to express my gratitude to Professors
  A. Poves and J. Luis Egido for their generous hospitality at the
  Department of Theoretical Physics of the
  {\em Universidad Autonoma de Madrid},
  where part of this work was accomplished.
  The Division de Physique Th\'eorique is an
  Unit\'e de Recherche des Universit\'es Paris XI et Paris VI, 
  associ\'ee au C.N.R.S.

%%%%%%%%%%%%%%%%%%%%%%%%%%%%%%%%%%%%%%%%%%%%%%%%%%%%%%%%%%%%%%%%%%%%%
%%%%%%%%%%%%%%%%%%%%%%%%%%%%%%%%%%%%%%%%%%%%%%%%%%%%%%%%%%%%%%%%%%%%%
%%%%%%%%%%%%%%%%%%%%%%%%%%%%%%%%%%%%%%%%%%%%%%%%%%%%%%%%%%%%%%%%%%%%%
%%%%%%%%%%%%%%%%%%%%      appendix        %%%%%%%%%%%%%%%%%%%%%%%%%%%
%%%%%%%%%%%%%%%%%%%%%%%%%%%%%%%%%%%%%%%%%%%%%%%%%%%%%%%%%%%%%%%%%%%%%
%%%%%%%%%%%%%%%%%%%%%%%%%%%%%%%%%%%%%%%%%%%%%%%%%%%%%%%%%%%%%%%%%%%%%
%%%%%%%%%%%%%%%%%%%%%%%%%%%%%%%%%%%%%%%%%%%%%%%%%%%%%%%%%%%%%%%%%%%%%
%%%%%%%%%%%%%%%%%%%%%%%%%%%%%%%%%%%%%%%%%%%%%%%%%%%%%%%%%%%%%%%%%%%%%

\begin{appendix}

\section{Eigenvectors and Eigenvalues}
\label{eigenapp}

  We follow Lage and de Almeida \cite{Lage_de_Almeida_82} in their
  stability analysis of the replica symmetric solution.
  The Hessian of the free energy is given schematically by:

  \[{\cal H} =
  \frac{\partial^2 f}{\partial [z_\alpha m_\alpha q_{\alpha\beta}]
  \partial [z_\alpha m_\alpha q_{\alpha\beta}]}
  = \left(\begin{array}{ccc}
  \begin{array}{cc} U(z,z) & V \\
                      & U  \end{array}  &
  \begin{array}{cc} W & X \\
                      & W  \end{array}  &
  \begin{array}{cc} Y & Z \\
                      & Y  \end{array}  \\
                                         &
  \begin{array}{cc} A(m,m) & B \\
                      & A  \end{array}  &
  \begin{array}{cc} C & D \\
                      & C  \end{array}  \\
                                         &
                                         &
  \begin{array}{ccc}  P(q,q) & Q & R \\
                        & P & Q \\
                        &   & P \end{array}
  \end{array}\right).\]

  The second equality defines the quantities $U, V, \ldots$
  with respect to their position in the Hessian. These quantities
  are the respective $G^{\alpha\beta}_{\epsilon\epsilon}$, etc \ldots as
  given in the main part. As will be seen later they can be expressed
  in terms of the different multispin correlation functions up to degree 4.
  The dimension of the Hessian is
  $n + n + \frac{n}{2}(n-1) = \frac{n}{2}(n+3)$ and
  equals the number of eigenvalues and eigenvectors to be found.
  In order to diagonalise the Hessian we construct the eigenvectors
  using the symmetry arguments exposed by
  de Almeida and Thouless \cite{de_Almeida_Thouless_78}.
  We do not construct a orthonormal set of eigenvectors,
  because the additional constraints destroy the symmetry with
  respect to replica permutations. Rather we content ourselves
  with three families of eigenvectors, each orthogonal to another,
  but not orthogonal within the families.

  We start with the
  eigenvector totally symmetric under permutations of the replicas:

  \[{\bf e}_{L} = \left\{ \begin{array}{r@{\quad:\quad}l}
        a & \mu = 1\ldots n \\
        b & \nu = 1\ldots n \\
        c & \alpha,\beta = 1\ldots \frac{n}{2}(n-1)\end{array}\right. \]

  According to conventional notation this vector is referred to as belonging to
  the longitudinal subspace; hence the subscript $L$. This is a
  subspace of dimension $d=3$, which is easily verified by constructing
  a orthogonal set of eigenvectors conserving the form prescribed above.
  The eigenvector equation
  ${\cal H} {\bf e}_{L} = \lambda_L {\bf e}_{L}$ reads

  \begin{eqnarray*}
  a [U + (n-1) V - \lambda_L] + b [W + (n-1) X]
                 + c [Y (n-1) + Z[\frac{n}{2}(n-1) - (n-1)]] &=& 0 \\
  a [W + (n-1) X] + b [A + (n-1) B - \lambda_L]
                 + c [C (n-1) + D[\frac{n}{2}(n-1) - (n-1)]] &=& 0 \\
  a [2Y + (n-2)Z] + b [2C + (n-2)D] + c [P + 2 (n-2) Q
                 + R [\frac{n}{2}(n-1) - 2 (n-1) -1] - \lambda_L] &=& 0
  \end{eqnarray*}

  The longitudinal eigenvector equation gives rise to a cubic characteristic
  equation for the eigenvalue $\lambda_L$:

  \[(\lambda^L)^3 + a (\lambda^L)^2 + b\lambda^L + c  = 0. \]

  We give its coefficients in the $n=0$ limit:

  \begin{eqnarray*}
  a &=& - [U-V] - [A-B] - [P-4\,Q+3\,R]
     = - \lambda_1^L - \lambda_2^L - \lambda_3^L\\
  b &=& - [W-X]^2 + [U-V] [P-4\,Q+3\,R] + 2 [Y-Z]^2 \\
     && + [A-B] [P-4\,Q+3\,R] + [U-V] [A-B] + 2[C-D]^2
      = \lambda_1^L \lambda_2^L
      + \lambda_2^L\lambda_3^L + \lambda_1^L\lambda_3^L  \\
  c &=& - [U-V] [A-B] [P-4\,Q+3\,R] - 2 [U-V] [C-D]^2 \\
     && + [W-X]^2 [P-4\,Q+3\,R]  + 4 [W-X] [Y-Z] [C-D] - 2 [Y-Z]^2 [A-B]
      = - \lambda_1^L \lambda_2^L \lambda_3^L.
  \end{eqnarray*}

  Define the following quantities:

  \begin{eqnarray*}
  \gamma &=& -\frac{a^2}{3} + b  \\
  \varrho &=& 2 \left(\frac{a}{3}\right)^3 - \frac{1}{3} ab + c \\
  \Gamma &=& \left(\frac{\gamma}{3}\right)^3
         + \left(\frac{\varrho}{2}\right)^2 \\
  \sigma^+ &=& \left(- \frac{\varrho}{2} + \sqrt{\Gamma} \right)^\frac{1}{3} \\
  \sigma^- &=& \left(- \frac{\varrho}{2} - \sqrt{\Gamma} \right)^\frac{1}{3},
  \end{eqnarray*}

  where only the real cubic roots are used. This allows us to
  write the three solutions as:

  \begin{eqnarray}\label{longitudinal_eigenvalues}
  \lambda_{L1} &=& \sigma^+ + \sigma^-             \nonumber \\
  \lambda_{L2} &=& -\frac{1}{2}(\sigma^+ + \sigma^-) +
                    \frac{i}{2}\sqrt{3}(\sigma^+ - \sigma^-)\\
  \lambda_{L3} &=& -\frac{1}{2}(\sigma^+ + \sigma^-) -
                    \frac{i}{2}\sqrt{3}(\sigma^+ - \sigma^-). \nonumber
  \end{eqnarray}

  The value of $\Gamma$ tells us whether the solutions are degenerate or not.

  \[\Gamma \left\{ \begin{array}{r@{\quad:\quad}l}
        < 0 & \mbox{distinct real solutions}\\
        = 0 & \mbox{degenerate real solutions}\\
        > 0 & \mbox{complex solutions}\\
    \end{array}\right. \]

  As has been recognized by Lage and de Almeida \cite{Lage_de_Almeida_82},
  occasionally $\Gamma$ is positive and so the eigenvalues become complex.
  This has been verified numerically by da Costa {\em et al.}
  \cite{da_Costa_Yokoi_Salinas_94}.

  The next eigenvectors, which will be called longitudinal anomalous, are
  constructed by breaking the symmetry of the longitudinal vector with
  respect to one replica, given by the distinct index $\theta$.

  \[{\bf e}_{LA} = \left\{ \begin{array}{r@{\quad:\quad}l}
        a & \mu = 1\ldots n, \mu \ne \theta \\
        g & \mu = \theta \\
        b & \nu = 1\ldots n, \mu \ne \theta \\
        c & \nu = \theta \\
        d & \alpha, \beta = 1\ldots \frac{n}{2}(n-1),
                            \alpha,\beta \ne \theta \\
        e & \alpha, \beta = \theta \\
        \end{array}\right. \]

  Consider the orthogonality condition for the first eigenvector,
  if $a=g$ or $b=c$.
  This condition results in trivial eigenvectors. In order to obtain
  non-trivial eigenvectors, the symmetry in both $n$-blocks has to be
  broken, giving $k n$ eigenvectors of the second family, where $k$ is
  the number of different choices of the parameters conserving the
  prescribed form. Orthogonality of the second family to the first
  family of eigenvectors requires:

  \begin{equation}
   g = (1-n) a \qquad
   c = (1-n) b \qquad
   e = (1-\frac{n}{2}) d ~.
  \end{equation}

  Therefore $k = 3$ and this choice gives rise to $3 n$ eigenvectors
  including the previous one. Writing down the characteristic equation
  for the longitudinal anomalous eigenvalues $\lambda_{LA}$ gives:

  \begin{eqnarray*}
  a [U - V - \lambda_{LA}] + b [W - X]
                 + d (\frac{n}{2}-1)[Y - Z] &=& 0 \\
  a [W - X] + b [A - B - \lambda_{LA}]
                 + d (\frac{n}{2}-1)[C- D] &=& 0 \\
  2 a [Y - Z] + 2 b [C - D]
         + d (\frac{n}{2}-1) [P + (n-4) Q + R (3-n) - \lambda_{LA}] &=& 0
  \end{eqnarray*}

  The longitudinal anomalous eigenvector equation also gives rise to a
  cubic characteristic equation for the eigenvalue $\lambda_L$:

  \[(\lambda^{LA})^3 + a (\lambda^{LA})^2 + b\lambda^{LA} + c  = 0. \]

  We quote its coefficients in the $n=0$ limit:

  \begin{eqnarray*}
  a &=& -U+V-A+B-P+4\,Q-3\,R  \\
  b &=& \left (W-X\right )^{2}-\left (U-V\right )\left (P-4\,Q+3\,R \right)
      + \left (2\,Y-2\,Z\right )\left (-Y+Z\right ) \\
   && - \left (A-B\right )\left (P-4\,Q+3\,R\right )
      - \left (U-V\right )\left (A-B\right )
      + \left (-C+D\right )\left (2\,C-2\,D\right ) \\
  c &=& \left (U-V\right )\left (A-B\right )\left (P-4\,Q+3\,R\right )
      - \left (U-V\right )\left (-C+D\right )\left (2\,C-2\,D\right ) \\
  &&  - \left (W-X\right )^{2}\left (P-4\,Q+3\,R\right )
      + \left (W-X\right )\left (-Y+Z\right )\left (2\,C-2\,D\right ) \\
  &&  + \left (2\,Y-2\,Z\right )\left (W-X\right )\left (-C+D\right )
      - \left (2\,Y-2\,Z\right )\left (-Y+Z\right )\left (A-B\right )
  \end{eqnarray*}

  Again the solutions may be written as in 
  equation \ref{longitudinal_eigenvalues},
  but now with modified coefficients $\gamma, \varrho, \ldots$.

  The eigenvalues will in general be different in each family, but since
  the characteristic equations for the {\em longitudinal} and the
  {\em anomalous longitudinal} become identical in $n=0$-limit, so
  do the eigenvalues, and then
  the {\em longitudinal} and {\em longitudinal anomalous}
  families collapse. There are $3 n$ eigenvectors and the three
  eigenvalues are each $n$-fold degenerate, including the three eigenvalues
  of the first eigenvector.

  Finally, there is the third family, called {\em replicon}.
  Breaking the symmetry with respect to permutations of pairs by
  distinguishing two indices $\theta$ and $\omega$, we obtain
  the following form for the eigenvectors:

  \[{\bf e}_{R} = \left\{ \begin{array}{r@{\quad:\quad}l}
        a & \mu = 1\ldots n, \mu \ne \theta,\omega \\
        g & \mu = \theta,\omega \\
        b & \nu = 1\ldots n, \mu \ne \theta, \omega \\
        c & \nu = \theta, \omega \\
        d & \alpha, \beta = 1\ldots \frac{n}{2}(n-1),
           \alpha,\beta \ne \theta \mbox{ and } \alpha,\beta \ne \omega\\
        e & \alpha, \beta = \theta \mbox{ or } \alpha, \beta = \omega \\
        f & (\alpha \beta) = (\theta\omega) \\
        \end{array}\right. \]

  Orthogonality of the third family to the previous two families
  of eigenvectors requires:

  \begin{equation}
  g = a \qquad
  c = b \qquad
  0 = f + (n-4) e + (3-n) d.
  \end{equation}

  The choice of equal non-vanishing entries would give the first family,
  so the recommended choice is

  \begin{equation}
  g = a = 0 \qquad
  c = b = 0 \qquad
  f = (2-n) e \qquad
  e = \frac{1}{2} (3-n) d.
  \end{equation}

  The characteristic equation is the solution for the replicon eigenvalue
  itself:

  \[\lambda_R = P - 2 Q + R.\]

  The replicon eigenvalue is independent of $n$ and identical in form
  to the result obtained by de Almeida and Thouless
  \cite{de_Almeida_Thouless_78}. This eigenvalue is
  $\frac{n}{2}(n-3)$-fold degenerate.

\subsection{The replica symmetric Ansatz}
\label{stabRS}

  In order to analyse the stability of the replica symmetric Ansatz, the
  quantities defined in the Hessian are to be evaluated
  assuming replica symmetry. This gives:

  \begin{eqnarray}\label{correlatorsI}
  A &=& \beta J_0 [1 - \beta J_0 (<s_\alpha^2> - <s_\alpha>^2)]
     =  \beta J_0 [1 - \beta J_0 (z - m^2)]  \nonumber \\
  B &=& \beta^2 J_0^2 [<s_\alpha>^2 - <s_\alpha s_\beta>]
     =  \beta^2 J_0^2 [m^2 - q]  \nonumber \\
  C &=& \beta^2 J^2 \beta J_0 [<s_\alpha><s_\alpha s_\beta>
          - <s_\alpha s_\beta^2>]
     =  \beta^2 J^2 \beta J_0 [m q - r]   \nonumber \\
  D &=& \beta^2 J^2 \beta J_0 [<s_\gamma><s_\alpha s_\beta>
     - <s_\alpha s_\beta s_\gamma>]
     =  \beta^2 J^2 \beta J_0 [m q - u]    \nonumber \\
  P &=& \beta^2 J^2 [1 - \beta^2 J^2 ( <s_\alpha^2 s_\beta^2>
     - <s_\alpha s_\beta>^2)]
     =  \beta^2 J^2 [1 - \beta^2 J^2 (s - q^2)]   \nonumber \\
  Q &=& \beta^4 J^4 [<s_\alpha s_\beta>^2 - <s_\alpha^2 s_\beta s_\gamma>]
     =  \beta^4 J^4 [q^2 - v]   \nonumber \\
  R &=& \beta^4 J^4 [<s_\alpha s_\beta>^2
     - <s_\alpha s_\beta s_\gamma s_\delta>]
     =  \beta^4 J^4 [q^2 - w] \\
  U &=& \beta \kappa [1 - \beta \kappa (<s_\alpha^2> - <s_\alpha^2>^2)]
     =  \beta \kappa [1 - \beta \kappa (z - z^2)]   \nonumber \\
  V &=& \beta^2 \kappa^2 [<s_\alpha^2>^2 - <s_\alpha^2 s_\beta^2>]
     =  \beta^2 \kappa^2 [z^2 - s]  \nonumber \\
  W &=& \beta J_0\beta \kappa [<s_\alpha> (<s_\alpha^2> - 1)]
     =  \beta J_0\beta \kappa [m (z - 1)]  \nonumber  \\
  X &=& \beta J_0\beta \kappa [<s_\alpha> <s_\alpha^2> - <s_\alpha s_\beta^2>]
     =  \beta J_0\beta \kappa [m z - r]  \nonumber  \\
  Y &=& \beta^2 J^2 \beta \kappa <s_\alpha s_\beta> [<s_\alpha^2>-1]
     =  \beta^2 J^2 \beta \kappa q [z-1]  \nonumber \\
  Z &=& \beta^2 J^2 \beta \kappa [<s_\alpha^2> <s_\alpha s_\beta>
     - <s_\alpha^2 s_\beta s_\gamma>]
     =  \beta^2 J^2 \beta \kappa [z q - v].  \nonumber
  \end{eqnarray}

  The second equality in each line results from the assumption of
  replica symmetry. Additional
  simplifications are due to the fact that for the $S=1$ spin model
  there are additional relations, e.g., $S^2 = S^4$, etc.
  Using the previously defined functions $\phi_k(y)$,
  see equation (\ref{phi_func}), the replica symmetric correlation
  functions introduced above may be explicitly written as:

  \begin{center}
  \begin{minipage}[t]{6cm}
  \begin{eqnarray*}
  m &=& <s_\alpha>   = \int {\cal D} \; y \phi_1(y) \\
  z &=& <s_\alpha^2> = \int {\cal D} \; y \phi_0(y) \\
  q &=& <s_\alpha s_\beta>     = \int {\cal D} \; y \phi_1(y)^2\\
  r &=& <s_\alpha s_\beta^2>   = \int {\cal D} \; y \phi_1(y)\phi_0(y)\\
  s &=& <s_\alpha^2 s_\beta^2> = \int {\cal D} \; y \phi_0(y)^2
  \end{eqnarray*}
  \end{minipage}
  \hskip 1cm
  \begin{minipage}[t]{6cm}
  \begin{eqnarray}\label{correlatorsII}
  t &=& <s_\alpha^3 s_\beta>
     = \int {\cal D} \; y \phi_1(y)^2 \nonumber \\
  u &=& <s_\alpha s_\beta s_\gamma>
     = \int {\cal D} \; y \phi_1(y)^3 \nonumber \\
  v &=& <s_\alpha^2 s_\beta s_\gamma>
     = \int {\cal D} \; y \phi_1(y)^2 \phi_0(y) \\
  w &=& <s_\alpha s_\beta s_\gamma s_\delta>
     = \int {\cal D} \; y \phi_1(y)^4. \nonumber
  \end{eqnarray}
  \end{minipage}
  \end{center}

\subsection{Paramagnetic phase}

  As is seen from the definitions of the multispin correlation
  functions (see equations \ref{correlatorsI} and \ref{correlatorsII})
  some of these vanish identically in the paramagnetic phase.
  Furthermore some of the combinations appearing in the calculation
  of the eigenvalues, e.g. $[W-X]= 0$, vanish too. This simplifies
  significantly the stability analysis in the paramagnetic phase,
  because the Hessian ${\cal H}$ and the matrix of the fourth
  derivatives become diagonal.

\end{appendix}

%%%%%%%%%%%%%%%%%%%%%%%%%%%%%%%%%%%%%%%%%%%%%%%%%%%%%%%%%%%%%%%%%%%%%
%%%%%%%%%%%%%%%%%%%%%%%%%%%%%%%%%%%%%%%%%%%%%%%%%%%%%%%%%%%%%%%%%%%%%
%%%%%%%%%%%%%%%%%%%%%%%%%%%%%%%%%%%%%%%%%%%%%%%%%%%%%%%%%%%%%%%%%%%%%
%%%%%%%%%%%%%%%%%%      bibliography        %%%%%%%%%%%%%%%%%%%%%%%%%
%%%%%%%%%%%%%%%%%%%%%%%%%%%%%%%%%%%%%%%%%%%%%%%%%%%%%%%%%%%%%%%%%%%%%
%%%%%%%%%%%%%%%%%%%%%%%%%%%%%%%%%%%%%%%%%%%%%%%%%%%%%%%%%%%%%%%%%%%%%
%%%%%%%%%%%%%%%%%%%%%%%%%%%%%%%%%%%%%%%%%%%%%%%%%%%%%%%%%%%%%%%%%%%%%
%%%%%%%%%%%%%%%%%%%%%%%%%%%%%%%%%%%%%%%%%%%%%%%%%%%%%%%%%%%%%%%%%%%%%

\clearpage
%\bibliographystyle{plain}
%\bibliography{litbank}

\newcommand{\noopsort}[1]{} \newcommand{\printfirst}[2]{#1}
  \newcommand{\singleletter}[1]{#1} \newcommand{\switchargs}[2]{#2#1}

%%%%%%%%%%%%%%%%%%%%%%%%%%%%%%%%%%%%%%%%%%%%%%%%%%%%%%%%%%%%%%%%%%%%%
%%%%%%%%%%%%%%%%%%%%%%%%%%%%%%%%%%%%%%%%%%%%%%%%%%%%%%%%%%%%%%%%%%%%%
%%%%%%%%%%%%%%%%%%%%%%%%%%%%%%%%%%%%%%%%%%%%%%%%%%%%%%%%%%%%%%%%%%%%%
%%%%%%%%%%%%%%%       figures captions        %%%%%%%%%%%%%%%%%%%%%%%
%%%%%%%%%%%%%%%%%%%%%%%%%%%%%%%%%%%%%%%%%%%%%%%%%%%%%%%%%%%%%%%%%%%%%
%%%%%%%%%%%%%%%%%%%%%%%%%%%%%%%%%%%%%%%%%%%%%%%%%%%%%%%%%%%%%%%%%%%%%
%%%%%%%%%%%%%%%%%%%%%%%%%%%%%%%%%%%%%%%%%%%%%%%%%%%%%%%%%%%%%%%%%%%%%
%%%%%%%%%%%%%%%%%%%%%%%%%%%%%%%%%%%%%%%%%%%%%%%%%%%%%%%%%%%%%%%%%%%%%

\end{document}